\documentclass[12pt]{iopart}
\usepackage[utf8]{inputenc}

\usepackage{hyperref}
\usepackage{siunitx}
\usepackage{graphicx}
\usepackage{xcolor}
\usepackage[subrefformat=parens]{subcaption}
\usepackage{cite}

\begin{document}
	\title{Airborne demonstration of a quantum key distribution receiver payload}
	
	\author{Christopher J. Pugh$^{1,2}$, Sarah Kaiser$^{1,2}$\footnote{Present Address: Department of Physics and Astronomy, Macquarie University, Balaclava Road, North Ryde, NSW, 2109, Australia}, Jean-Philippe Bourgoin$^{1,2}$, Jeongwan Jin$^{1,2}$, Nigar Sultana$^{1,3}$, Sascha Agne$^{1,2}$, Elena Anisimova$^{1,2}$, Vadim Makarov$^{2,1,3}$, Eric Choi$^{1}$\footnote{Present Address: Magellan Aerospace, 3701 Carling Avenue, Ottawa, Ontario K2H~8S2, Canada}, Brendon L. Higgins$^{1,2}$ and Thomas Jennewein$^{1,2}$}
	\address{$^1$ Institute for Quantum Computing, University of Waterloo, 200 University Avenue West, Waterloo, Ontario N2L~3G1, Canada}
	\address{$^2$ Department of Physics and Astronomy, University of Waterloo, 200 University Avenue West, Waterloo, Ontario N2L~3G1, Canada}
	\address{$^3$ Department of Electrical and Computer Engineering, University of Waterloo, 200 University Avenue West, Waterloo, Ontario N2L~3G1, Canada}
	\ead{thomas.jennewein@uwaterloo.ca}

	\begin{abstract}
		Satellite-based quantum terminals are a feasible way to extend the reach of quantum communication protocols such as quantum key distribution (QKD) to the global scale. To that end, prior demonstrations have shown QKD transmissions from airborne platforms to receivers on ground, but none have shown QKD transmissions from ground to a moving aircraft, the latter scenario having simplicity and flexibility advantages for a hypothetical satellite. Here we demonstrate QKD from a ground transmitter to a receiver prototype mounted on an airplane in flight. We have specifically designed our receiver prototype to consist of many components that are compatible with the environment and resource constraints of a satellite. Coupled with our relocatable ground station system, optical links with distances of \SIrange[range-units=single,range-phrase=--]{3}{10}{\km} were maintained and quantum signals transmitted while traversing angular rates similar to those observed of low-Earth-orbit satellites. For some passes of the aircraft over the ground station, links were established within \SI{10}{\s} of position data transmission, and with link times of a few minutes and received quantum bit error rates typically \SIrange[range-units=single,range-phrase=--]{{\approx}3}{5}{\percent}, we generated secure keys up to \SI{868}{\kilo b} in length. By successfully generating secure keys over several different pass configurations, we demonstrate the viability of technology that constitutes a quantum receiver satellite payload and provide a blueprint for future satellite missions to build upon.
		
	\end{abstract}
	
	\noindent{\it Keywords\/}: quantum key distribution, quantum communication, quantum cryptography, quantum optics, photon detection, quantum satellite payload
	\maketitle

	\section{Introduction}
	
	Quantum key distribution (QKD)~\cite{Bennet84,SBC09} establishes cryptographic keys between two distant parties in a way that is cryptanalytically unbreakable. Ground based implementations of QKD using optical fiber links are limited to distances of a few hundred kilometers due to absorption losses, which scale exponentially with distance, leading to insufficient signal-to-noise~\cite{SWV09,LCW10,KLH15}. Alternatively, free space links have been demonstrated over ground with varying distances, both in stationary~\cite{BHK98,HND02,UTSComm07,SWF07,VDS14} and moving~\cite{Nauerth13,WYLZ13,BHG15} configurations. But despite losses due to geometric effects scaling quadratically with distance, the addition of atmospheric absorption and turbulence, and the necessity of having clear line of sight, limit terrestrial free-space transmissions to, also, a few hundred kilometers.

	Much greater distances could be spanned in free-space transmissions outside Earth's atmosphere. Utilizing orbiting satellites therefore has potential to allow the establishment of global QKD networks, with ``quantum'' satellites acting as intermediaries. Such satellites could operate as untrusted nodes linking two ground stations simultaneously~\cite{HBK00,TCS14}, or trusted nodes connecting any two ground stations on Earth at different times~\cite{RTG02,UJK09,VJT08,EAW11,MYM11,YCL13,VDT16}. The majority of such analyses propose a quantum downlink, where photons are generated at the satellite and transmitted to receivers on the ground. Since 2010, the Canadian Space Agency (CSA) has studied the proposed Quantum Encryption and Science Satellite (QEYSSat)~\cite{JBH14}, where a mission concept was developed in partnership with COM~DEV (now Honeywell Aerospace). This concept, in contrast to many other missions, proposes a quantum uplink, placing the receiver on the satellite while keeping the quantum source at the ground station.

	Under similar conditions, the uplink configuration has a lower key generation rate than the downlink, owing to atmospheric turbulence affecting the beam path earlier in the propagation. Nevertheless, comprehensive theoretical comparative study of QKD under uplink and downlink conditions---which included the effects of atmospheric turbulence, absorption, beam propagation, optical component losses, detector characteristics, noise contributions, and representative pointing and collection capabilities at a hypothetical satellite---concluded that an uplink approach is viable, with the reduction in generated key bits (compared to downlink) being less than one order of magnitude~\cite{BMH13}. Importantly, an uplink also possesses a number of advantages over a downlink, including relative simplicity of the satellite design, not requiring high-rate true random number generators, relaxed requirements on data processing and storage (only the photon reception events need be considered, which are many orders of magnitude fewer than the source events), and the flexibility of being able to incorporate and explore various different quantum source types with the same receiver apparatus (which would have major associated costs were the source located on the satellite, as for downlink). Recently, China launched a quantum science satellite which aims to perform many quantum experiments with optical links between space and ground~\cite{G16,C16}. However its exact capabilities are unverified as no details or results have been published at this time.

	Demonstrations of QKD with moving and airborne platforms take important steps to verifying the readiness of quantum technology, and the supporting classical technology, for deployment within a satellite payload. To date, however, reported demonstrations of QKD with aircraft have operated exclusively in the downlink configuration~\cite{Nauerth13,WYLZ13}, where the quantum states are generated and transmitted from the aircraft to a receiver at a stationary ground location. Here we demonstrate QKD uplink to a receiver on a moving aircraft. Our apparatuses incorporate coarse- and fine-pointing systems necessary to establish and maintain optical link, quantum source and measurement components that conduct polarization-encoded QKD, and suitable post-processing algorithms to extract secure key. The results show good performance at the same angular rates exhibited by low-Earth-orbit (LEO) satellites.

	Our QKD receiver makes extensive use of components custom-designed according to the mass, volume, power, thermal, and vacuum operating environment requirements of systems to be embedded in a satellite payload---many components are already space suitable, and others have a clear path to flight. In a recent study, conducted with the University of Toronto Institute for Aerospace Studies Space Flight Laboratory (UTIAS SFL), a realistic satellite concept was developed, incorporating the space-ready receiver apparatus demonstrated here, integrated into the flight-proven NEMO-150~\cite{SFL} micro-satellite bus (see below). This, together with our airborne operational demonstration, illustrates the technological advancements made towards the development of a space-suitable QKD receiver, and highlights the feasibility and technological readiness of an uplink QKD satellite.

	\section{Apparatuses and Methods}

	\subsection{Concept}

	\begin{figure}[th]
		\centering
		\includegraphics[width=0.8\linewidth]{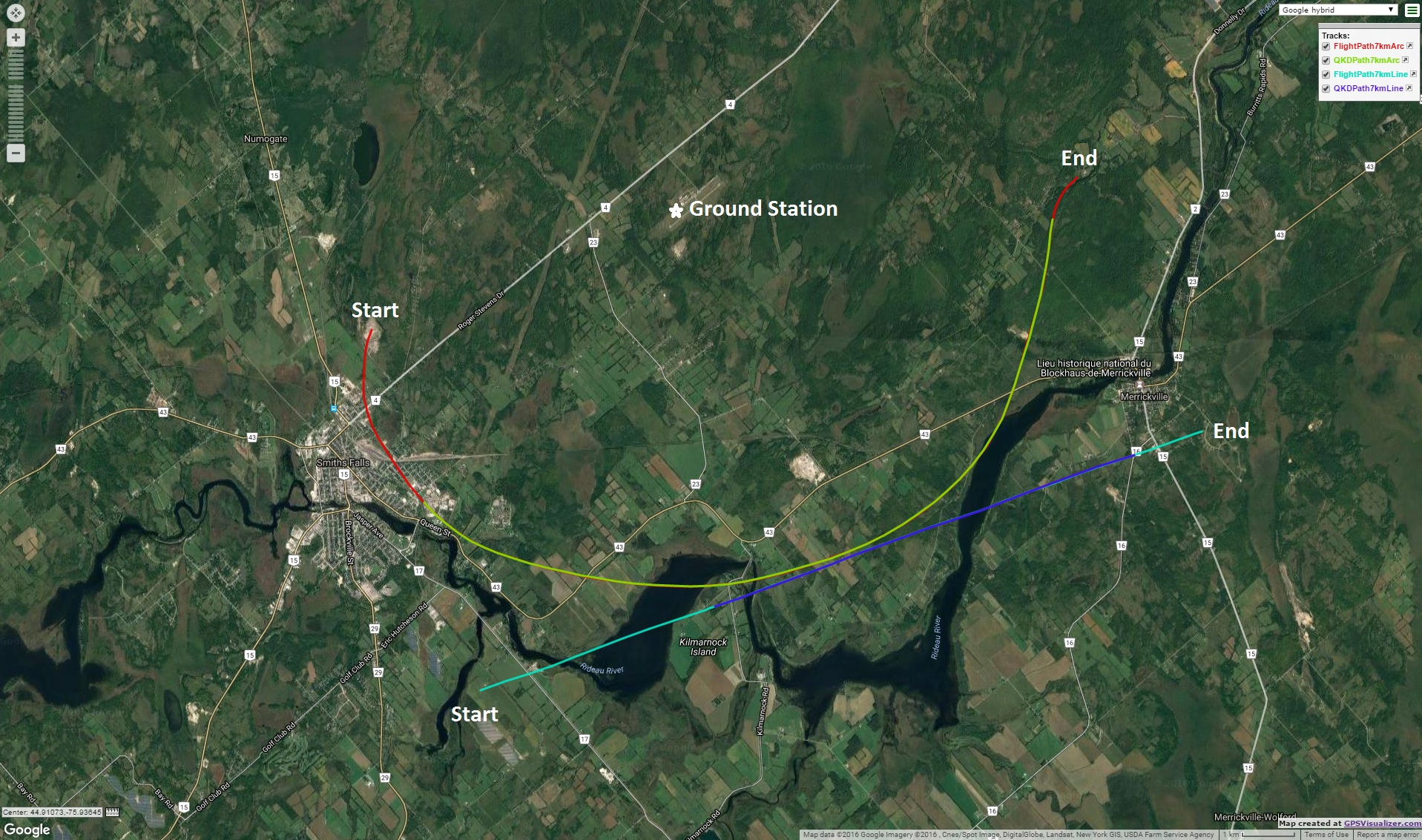}
		\caption{Flight paths for the \SI{7}{\km} arc and line, followed from left to right. The star indicates the location of the ground station at Smith Falls--Montague Airport. The inner portions represent where the quantum link was active. Photo produced using GPSVisualizer.com, map data \textcopyright{} 2016 Google, imagery \textcopyright{} 2016 Cnes/Spot Image, DigitalGlobe, Landsat, New York GIS, USDA Farm Service Agency.}
		\label{fig:Flight7km}
	\end{figure}

	The apparatuses for our demonstration consist of a QKD source and transmitter, located at a ground station near the airstrip of Smiths Falls--Montague Airport, and a QKD receiver, located on a Twin Otter research aircraft from the National Research Council of Canada. Optical links were only attempted at night, to limit optical noise. One systems-test day-time flight was conducted (where the optical links were not attempted), followed by night-time flights.

	Two night-time flights were conducted, each two hours duration and consisting of several passes of varying trajectories. Optical links were established using tracking feedback to two-axis motors, guided by strong beacon lasers (at a wavelength different from the quantum signal) and an imaging camera, at each of the two sites. The QKD signals produced by the source were guided through a telescope on the ground and pointed to the receiver on the aircraft. There the QKD signal polarizations and times-of-arrival were recorded for later correlation and processing to complete the QKD protocol and extract the key.
	
	We focused on two path types: arcs with (approximately) constant radius around the ground station, and straight lines past the ground station. For straight line paths, the distance we quote is the minimum. Over the two nights we performed 14 passes with nominal distances of \SI{3}{\km}, \SI{5}{\km}, \SI{7}{\km}, and \SI{10}{\km}, in both line and arc configurations, at an altitude of \SI{\approx 1.6}{km} above sea level---see, for example, Fig.~\ref{fig:Flight7km} for the flight path of an arc at \SI{7}{\km} radius. For this flight mission concept, a sequence of GPS coordinates was calculated for each flight, with the start angle relative to the ground station and the distance were used as input. These coordinates were transferred to the flight software of the aircraft by the pilots. We developed a decision tree such that, based on the observed performance of each pass, we could immediately select an appropriate course of action (e.g. to perform troubleshooting or collect data under different conditions). That a mission concept such as this is viable shows that a similar mission concept, appropriate for an orbiting satellite receiver, can realistically be achieved.
	
	\subsection{Source and Transmitter}

	\begin{figure}[th]
		\centering
		\begin{subfigure}[c]{0.475\linewidth}
			\centering\includegraphics[page=1,width=\linewidth]{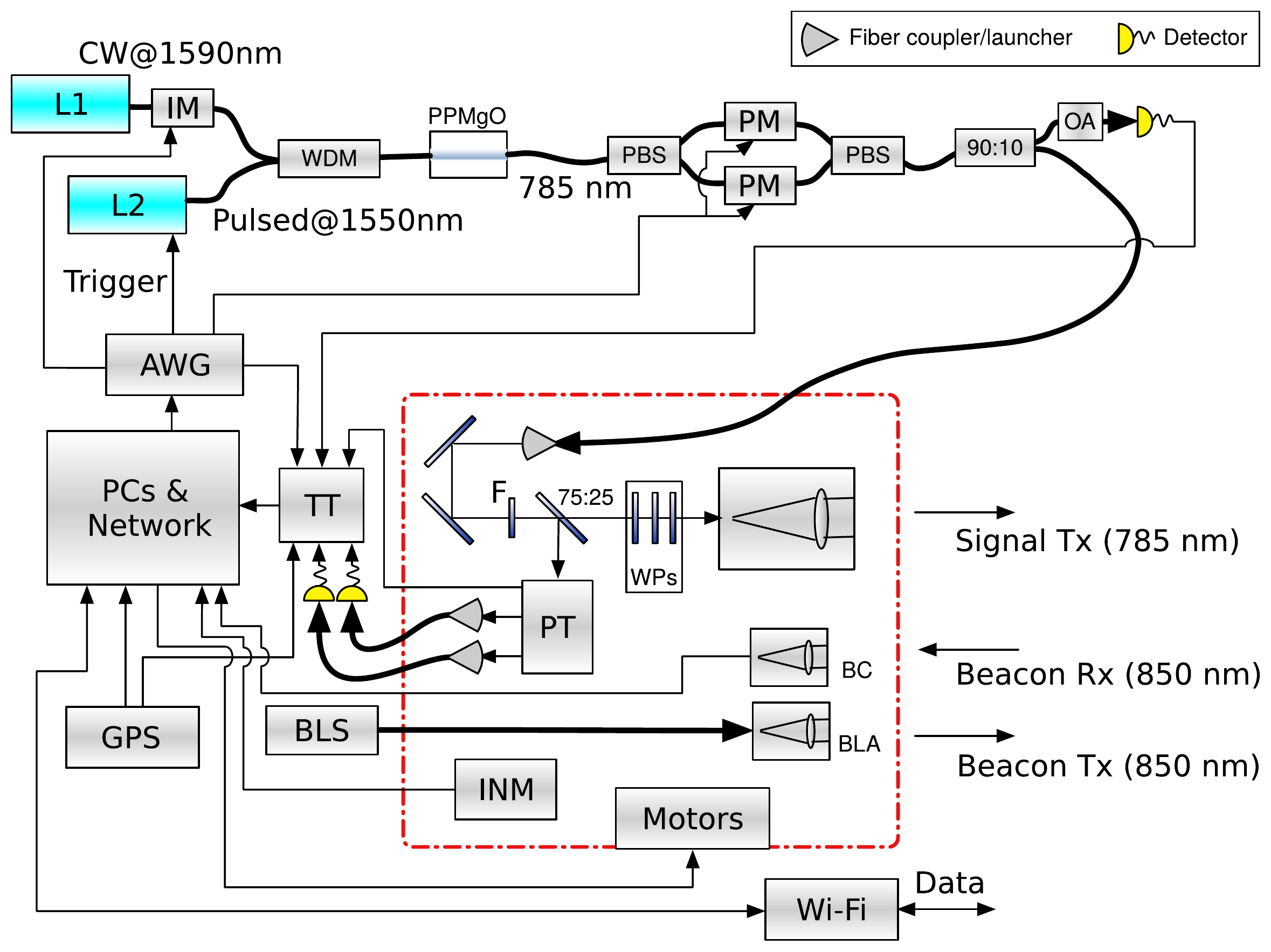}
		\end{subfigure}
		\hspace{0.02\linewidth}
		\begin{subfigure}[c]{0.475\linewidth}
			\centering\includegraphics[width=\linewidth]{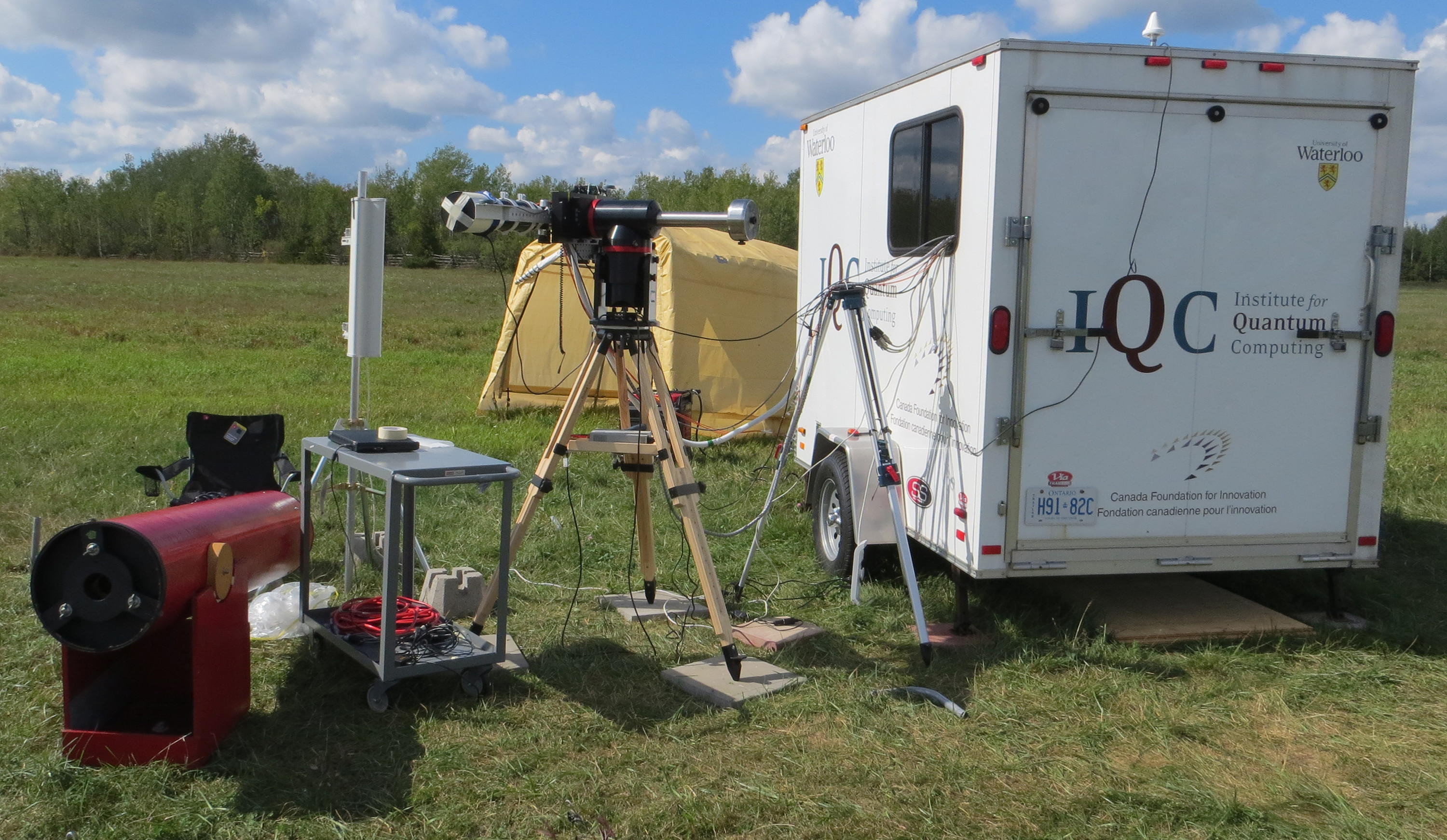}
		\end{subfigure}
		\caption{Left, schematic diagram of the quantum source and transmitter apparatus. AWG, arbitrary waveform generator; WDM, wavelength division multiplexer; PBS, polarizing beam splitter; OA, optical attenuator; F, band-pass filter; PT, polarization tomography; TT, time tagger. Other acronyms and details given in the text. The red border indicates components that are mounted on the motors. Right, ground station located at Smiths Falls--Montague airport, showing (right-to-left) the trailer where the source is located, motor mount with transmitter telescope attached, Wi-Fi antenna, and calibration telescope.}
		\label{fig:Transmitter}
	\end{figure}

	Our QKD source is a significantly improved version of a previous-generation apparatus~\cite{YMB13}, implementing BB84 with decoy states~\cite{LMC05} at \SI{400}{\mega\Hz}. Weak coherent pulses at \SI{785}{\nm} wavelength are generated by combining a narrow-band \SI{1590}{\nm} continuous-wave (CW) laser (L1) with \SI{1550}{\nm} triggered-pulsing laser (L2) through sum frequency generation in a periodically poled magnesium oxide (PPMgO) waveguide (see Fig.~\ref{fig:Transmitter}). For each pulse, one of three intensity levels is chosen: signal, decoy, or vacuum, with probabilities of \SI{80}{\percent}, \SI{14}{\percent}, and \SI{6}{\percent}, respectively. Signal and decoy levels are generated using a fast electro-optical intensity modulator (IM) calibrated to emit $\mu \approx 0.5$ and $\nu \approx 0.1$ mean photon number at the entrance of the transmitter telescope, respectively. The vacuum state is made by suppressing the laser trigger.

	Each of the four BB84 polarizations---horizontal (H), vertical (V), diagonal (D), and anti-diagonal (A)---are imposed using two electro-optical phase modulators (PMs), each in one arm of a balanced Mach-Zehnder polarization interferometer. With a balanced input (D), the PMs can address any point on the circle through D, right-circular (R), A, and left-circular (L). A subsequent unitary rotation takes these to D, H, A, and V, respectively. The intensity and polarization states are generated according to a randomized sequence that repeats every 1000 pulses. Although this is insecure, it is sufficient for our demonstration, while upgrading to a fully random sequence (e.g., given by a quantum random number generator) is straightforward and a suitable system has been identified.

	Pulse intensities are measured locally through the weak output of an optical fiber splitter (90:10) connected to a silicon avalanche photodiode (Si-APD) operating in Geiger mode with active quenching. The bulk of the pulse power is guided from the source to the transmitter through single-mode optical fiber. The beam passes through a \SI{785}{\nm} band-pass (\SI{3}{\nm} bandwidth) filter (to impede Trojan-horse attacks~\cite{JAK14}) and then a 75:25 beam splitter. We employ a polarization correction system to undo the unknown unitary rotation applied by the single-mode fiber---the reflected \SI{25}{\percent} of pulses undergo characterization, while the remaining \SI{75}{\percent} of pulses pass through a triplet of wave plates (WPs) in motorized rotation stages that apply a compensation operation to the states, and are finally transmitted through a \SI{12}{\cm} diameter Sky-Watcher BK 1206AZ3 refractive telescope.

	The polarization characterization subsystem consists of two beam paths, where each path passes through a port of a rotating chopper wheel that contains linear polarizers. The linear polarizers are each calibrated to project onto the H, V, D, or A state---however, one of the two beam paths contains a quarter-wave plate just prior to the chopper wheel, thereby facilitating projections onto a tomographically complete set of three polarization bases: H/V, D/A, and R/L. The actual state any given photon is projected to depends on which blade of the wheel is open at the time the photon passes through (the rotation of the wheel is also recorded).

	The two beams are each coupled into fiber and directed to Si-APDs. With near real-time analysis of source and detection data (performed on per-second integrated counts), we obtain tomographic reconstructions, for each of the generated polarization states, of the states at the transmitter after the rotation applied by the fiber. We then optimize the compensating wave plate triplet (a sequence of quarter-, half-, and quarter-wave plate) to maximize the fidelity of the states expected after compensation with the nominally generated states. The optimal positions are given to the motorized stages, applying the (rotated) wave plates to pulses that are then transmitted through the telescope towards the receiver.

	During our airborne trials, the QKD source optics and electronics, as well as computers for data recording and pointing feedback, were located inside of a trailer to maintain thermal and humidity stability. The transmitter pointing stages, polarization characterization optics, and telescope were located just outside the trailer, with cabling running through a small window. Equipped with an electric generator, our ground station is relocatable and self-sufficient. 
	
	\subsection{Receiver}

	\begin{figure}[th]
		\centering
		\begin{subfigure}[c]{0.475\linewidth}
			\centering\includegraphics[width=\linewidth]{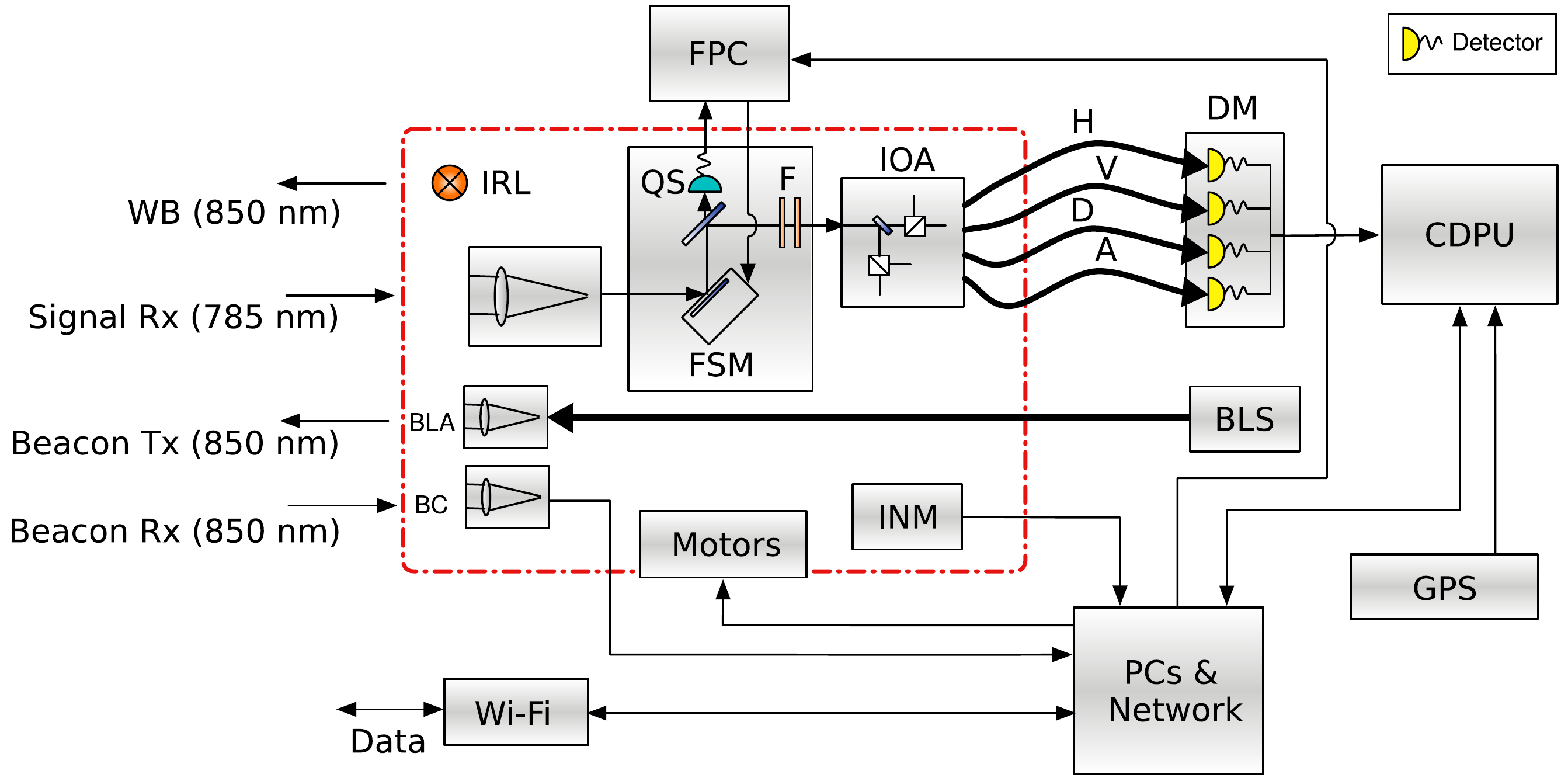}
		\end{subfigure}
		\hspace{0.02\linewidth}
		\begin{subfigure}[c]{0.475\linewidth}
			\centering\includegraphics[width=\linewidth]{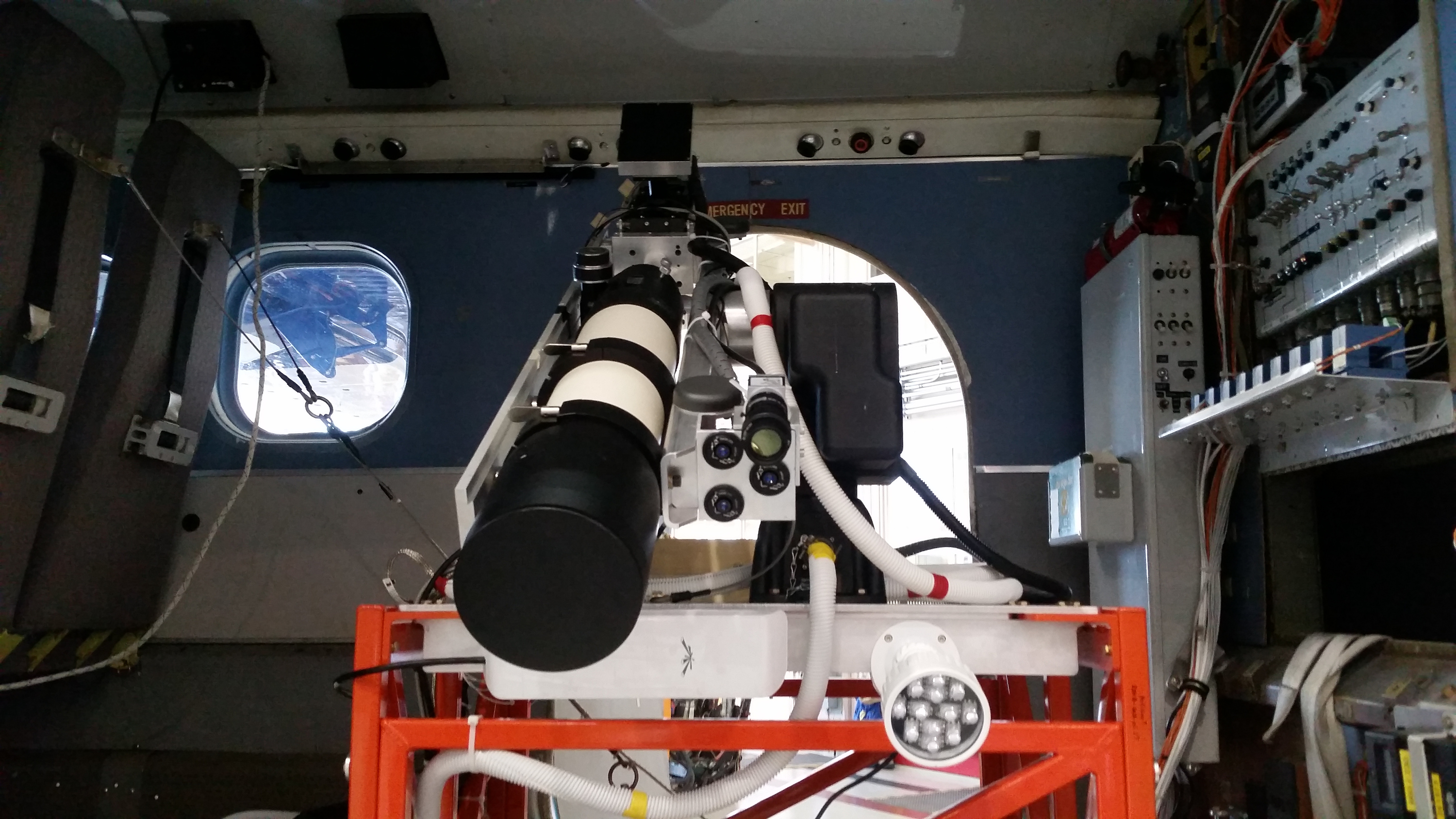}
		\end{subfigure}
		\caption{Left, schematic diagram of the receiver apparatus. F, band-pass filters; WB, wide-field beacon (produced by the IRL). Other acronyms and details given in the text.  The red border indicates components that are mounted on the motors. Right, receiver apparatus facing out the port-side door of the NRC Twin Otter research aircraft, showing (clockwise) the telescope, beacon assembly, motor mount, IRL, Wi-Fi antenna, and FPC (behind). Other components not visible (primarily electronics) are mounted in front of the seats seen at the left of the picture.}
		\label{fig:Receiver}
	\end{figure}

	At the receiver (Fig.~\ref{fig:Receiver}), the signal is collected by a Tele Vue NP101is refractive telescope with a \SI{10}{\cm} aperture, and coupled into a sequence of custom components developed under contract with the Canadian Space Agency~\cite{CSA_QKDR}. First of these is a fine-pointing unit (FPU), developed with Institut National d'Optique (INO) and Neptec Design Group, which guides both the quantum and beacon signals with a fast-steering mirror (FSM). Inside the FPU, a dichroic mirror separates the quantum and beacon signals---the beacon is reflected towards a quad-cell photo-sensor (QS), providing position feedback to a fine-pointing controller (FPC) that guides the fast-steering mirror in a closed loop~\cite{PKB17}. The FPU, measuring \SI{330 x 240 x 95}{\mm} and \SI{2.61}{\kg}, has a \SI{\pm 0.3}{\degree} field of view.
	
	The collected quantum beam is guided through a \SI{50}{\um} pinhole, acting as a spatial-mode filter~\cite{SCB15}, followed by a pair of \SI{785}{\nm} (\SI{3}{\nm} bandwidth) spectral filters. It then passes into a custom integrated optical assembly (IOA) developed with INO, containing a passive-basis-choice polarization analysis module with a 50:50 beam splitter and polarizing beam splitters. The IOA, measuring \SI{48.2 x 56.8 x 120}{\mm} and \SI{129}{\g}, produces four beams coupled into multimode fibers, corresponding to the four BB84 measurement states (H, V, D, and A) with state contrasts between 532:1 and 2577:1.

	The four IOA output fibers are guided to a detector module (DM) containing four Excelitas Technologies SLiK Si-APD detectors operating in Geiger mode with passive quenching. The DM measures \SI{30 x 127 x 143}{\mm} and \SI{516}{\g}, and operates at \SI{2.3}{\W} steady state (including thermoelectric cooling of detector active areas to \SI{-20}{\celsius}) to give a detection efficiency of \SI{{\approx}45}{\percent}, biased \SI{28}{\V} above breakdown.

	The detectors trigger low-voltage differential signalling pulses which are measured at a control and data processing unit (CDPU) based on Xiphos Systems Corporation's Q7 processor card (recently flown on GHGSat~\cite{GHG}) with a custom daughterboard. The CDPU utilizes an ARM Cortex-A9 processor and measures \SI{25 x 107 x 118}{\mm}, \SI{129}{\g}, drawing \SI{4.5}{\W} while operating. A field-programmable gate array embedded in the CDPU is programmed to implement time-tagging of detection pulses with a resolution of \SI{78}{\ps}, while data storage, communication, and processing software running in the Linux operating system implement the receiver-side QKD protocol.
	
	The receiver telescope was mounted facing out the cabin door on the port side of the aircraft, and flown with the door removed. The electronics and computers were located six feet forward in the aircraft cabin, and optical fibers and cables conducted signals between the electronics and the receiver telescope and pointing equipment.

	\subsection{Acquisition and Calibration}
	
	The transmitter and receiver each have a beacon laser assembly (BLA) consisting of three fiber launchers with fixed divergence angles of \SI{0.74}{\degree} and individual tip/tilt control. These are mounted on each telescope and fed strong (\SI{{\approx}40}{\mW}) \SI{850}{\nm} laser light from fiber-coupled beacon laser source (BLS) arrays located away from the telescopes. A beacon camera (BC)---a 50 frame-per-second, 2 megapixel imaging camera with an \SI{850}{\nm} band-pass filter (\SI{10}{\nm} bandwidth)---is also mounted to each telescope.

	Each telescope is attached to a commercial two-axis motor system (transmitter: ASA DDM85 Standard, receiver: FLIR PTU-D300E), providing first-stage ``coarse'' pointing. When light at the beacon wavelength is visible as a bright spot on the camera image, our custom pointing software (running on PCs at each site) controls the angular speeds of the motors to minimize the deviation of the spot's position from a calibrated reference position. The control feedback loop incorporates the estimated angular speed of the spot (based on position differences between recent images, taking into account previous motor motions), a factor proportional to the spot's current deviation, and a factor proportional to the accumulated (integrated) spot deviations. The pointing software operates as a state machine, and also includes a ``coasting'' state to handle short drop-outs of the beacon signal, and ``acquiring'' and ``searching'' states to support the initial acquisition of the beacon.
	
	To achieve initial acquisition, we employ inertial navigation modules (INMs), containing GPS receivers and attitude sensors, mounted to the telescopes. Each site transmits their GPS location to the other site via a classical RF (Wi-Fi) link, and then calculates the other site's orientation relative to its own based on its local attitude data.  During initial testing, the INMs exhibited an attitude uncertainty of about \SI{\pm2.5}{\degree}---significantly larger than the \SI{0.74}{\degree} divergence of the beacon lasers. To mitigate this, we turn on a bright infra-red light-emitting diode array (IRL) at the receiver with much greater divergence (of order \SI{80}{\degree}), allowing the transmitter to find, and point towards, the receiver. Once the receiver sees the transmitter's beacon spot in its camera image and has moved to position, the IRL is switched off, and two-way beacon tracking continues for the remainder of the pass.

	A necessary practical feature of our transmitter and receiver apparatuses is that they can be independently calibrated, as they would not be co-located prior to establishing a link (much like for a satellite mission). To align each of the beacon lasers with the quantum signal beam path, we first inject alignment laser power into each telescope, and point the telescope towards a separate larger-diameter (\SI{{\approx}20}{\cm}) telescope, located \SI{{\approx}20}{\m} away, equipped with a camera imaging the far field. We then observe the position of the beacon beams on the camera image, and adjust the tip and tilt of each beacon fiber launcher to center its output over the signal spot. To calibrate the reference position of the beacon camera at the transmitter and the collimation of the transmitted quantum beam, we optimize the power received (using the alignment laser injected into the transmitter telescope) at another telescope located at a sufficient distance \SI{{\approx}850}{\m} down the runway. The receiver beacon camera, which has greater tolerance due to the receiver's fine-pointing unit, is calibrated using a corner cube located \SI{\approx 50}{\m} away in the NRC hangar. These alignments were done prior to each flight. These independent calibrations allowed link acquisition to begin immediately upon the arrival of the airplane in the vicinity of the ground station.

	\section{Results}
	
	In total, seven of the 14 airplane passes over the ground station successfully established a quantum signal link. Issues, including minor equipment failures (e.g., a loose beacon camera lens) and accidental controller misconfigurations, particularly hampered link establishment during the first night---two of the seven attempts were successful. These issues were addressed during the intervening day, and the second night had considerably better link establishment rate---five of seven attempts. (We attribute the two failures on the second night---both attempted straight-line paths---to the fixed orientation of the Wi-Fi transceiver at the aircraft being poor for this geometry, particularly at the beginning of a pass.)

	Secret key was extracted out of six of the seven successful passes. From data collected during these passes, we observe the performance of the system at various distances and with angular speeds. Circular-arc passes allowed us to demonstrate longer duration of key exchange, compared to straight-line passes, as the receiver telescope held a relatively constant position during the pass, making link establishment and pointing easier. Straight line passes, however, are much more representative of a satellite passing over a ground station, as they simulate the change in angular speed that would be experienced during such a pass. The maximum angular rate is reached when the airplane is closest to the ground station for that pass---the greatest maximum angular rate we measured for our passes was \SI{1.28}{\degree/\s} at a distance of \SI{3}{\km} (arc). This angular rate is consistent with overflying LEO spacecraft such as \SI{0.72}{\degree/\s} for a \SI{600}{\km} orbit, as baselined for QEYSSat, or \SI{1.2}{\degree/\s} for the International Space Station (ISS).
	
	\begin{table}
		\centering
		\caption{Summary of data from passes where a quantum link was established. All times are UTC. Except where indicated (*), secure key lengths incorporate finite-size effects.}
		\tiny
		\begin{tabular}{l|l|l||l|l|l|l|l}
			\hline\hline
			\multicolumn{1}{r|}{Pass} & \SI{5}{\km} & \SI{7}{\km} & \SI{5}{\km} & \SI{3}{\km} & \SI{3}{\km} & \SI{7}{\km} & \SI{10}{\km} \\
			& arc 1 & line & arc 2 & line & arc & arc & arc \\
			& 2016-09-21 & 2016-09-21 & 2016-09-22 & 2016-09-22 & 2016-09-22 & 2016-09-22 & 2016-09-22 \\
			Parameter & 2:57:45 & 3:30:45 & 1:15:23 & 2:19:33 & 2:24:45 & 2:42:16 & 2:57:42 \\
			\hline
			Classical link duration [\si{\s}] & 288 & 172 & 352 & 34 & 170 & 210 & 289 \\
			Quantum link duration [\si{\s}] & 235 & 158 & 250 & 33 & 158 & 206 & 269 \\
			Mean speed [\si{\km/\hour}] & 208 & 200 & 198 & 236 & 216 & 259 & 212 \\
			Maximum angular speed [\si{\degree}] & 0.76 & 0.45 & 0.75 & 1.0 & 1.28 & 0.60 & 0.37 \\
			Transmitter pointing error ($10^{-3}$)[\si{\degree}] & 22.0 & 4.85 & 1.33 & 3.42 & 2.91 & 1.58 & 2.82 \\
			Receiver pointing error ($10^{-3}$)[\si{\degree}] & 125 & 126 & 63.0 & 86.5 & 89.8 & 78.6 & 87.2 \\
			Receiver fine-pointing error ($10^{-3}$)[\si{\degree}] & 2.73 & 9.98 & No data & 2.62 & 2.39 & 3.01 & 12.7 \\
			Source QBER [\si{\percent}] & 5.08 & 3.58 & 3.32 & 2.66 & 4.37 & 2.80 & 3.39 \\
			Signal QBER [\si{\percent}] & 13.13 & 5.24 & 3.42 & 2.96 & 5.20 & 2.96 & 3.30 \\
			Decoy QBER [\si{\percent}] & 19.54 & 11.1 & 6.13 & 6.35 & 7.93 & 5.97 & 8.46 \\
			Theoretical loss [\si{\dB}] & 52.1 & 41.6--44.8 & 28.1 & 33.3--35.1 & 30.9 & 32.1 & 39.9 \\
			Mean measured loss [\si{\dB}] & 48.0 & 51.1 & 34.5 & 39.5 & 34.4 & 39.4 & 42.6 \\
			Error correction efficiency & 1.4 & 1.16 & 1.33 & 1.4 & 1.18 & 1.46 & 1.27 \\
			Signal-to-noise threshold & 0 & 1500 & 2000 & 1000 & 1000 & 2000 & 2500 \\
			Sifted key length [bits] & 152508 & 95710 & 5212446 & 853066 & 5102122 & 2348086 & 1175317 \\
			Secure key length [bits] & None & 9566* & 867771 & 71648 & 44244 & 200297 & 70947 \\
			\hline\hline
		\end{tabular}
		\label{tab:Flights}
	\end{table}
	
	Table~\ref{tab:Flights} summarizes the seven passes where quantum signal was successfully transmitted to the receiver aboard the aircraft. Passes typically lasted a few minutes, with the aircraft travelling at \SIrange{198}{259}{\km/\hour}. To quantify pointing performance, we define the typical pointing error as the measured distance of the beacon spot from the calibrated reference point on the camera image, discarding times when the motors had just begun tracking. The mean typical pointing error at the transmitter varied from \SIrange{0.00133}{0.0220}{\degree} over the passes; at the receiver, it was \SIrange{0.0630}{0.126}{\degree}. The receiver's fine-pointing unit measured pointing errors similar to the pointing error of the transmitter, between \SIrange{0.00239}{0.0127}{\degree}, where the deviation was measured from the centre of the quad-cell sensor. (These values are used in the link analysis model, below.)
	
	\begin{figure}[thp]
		\centering
		\begin{subfigure}[c]{0.475\linewidth}
			\centering\includegraphics[width=\linewidth]{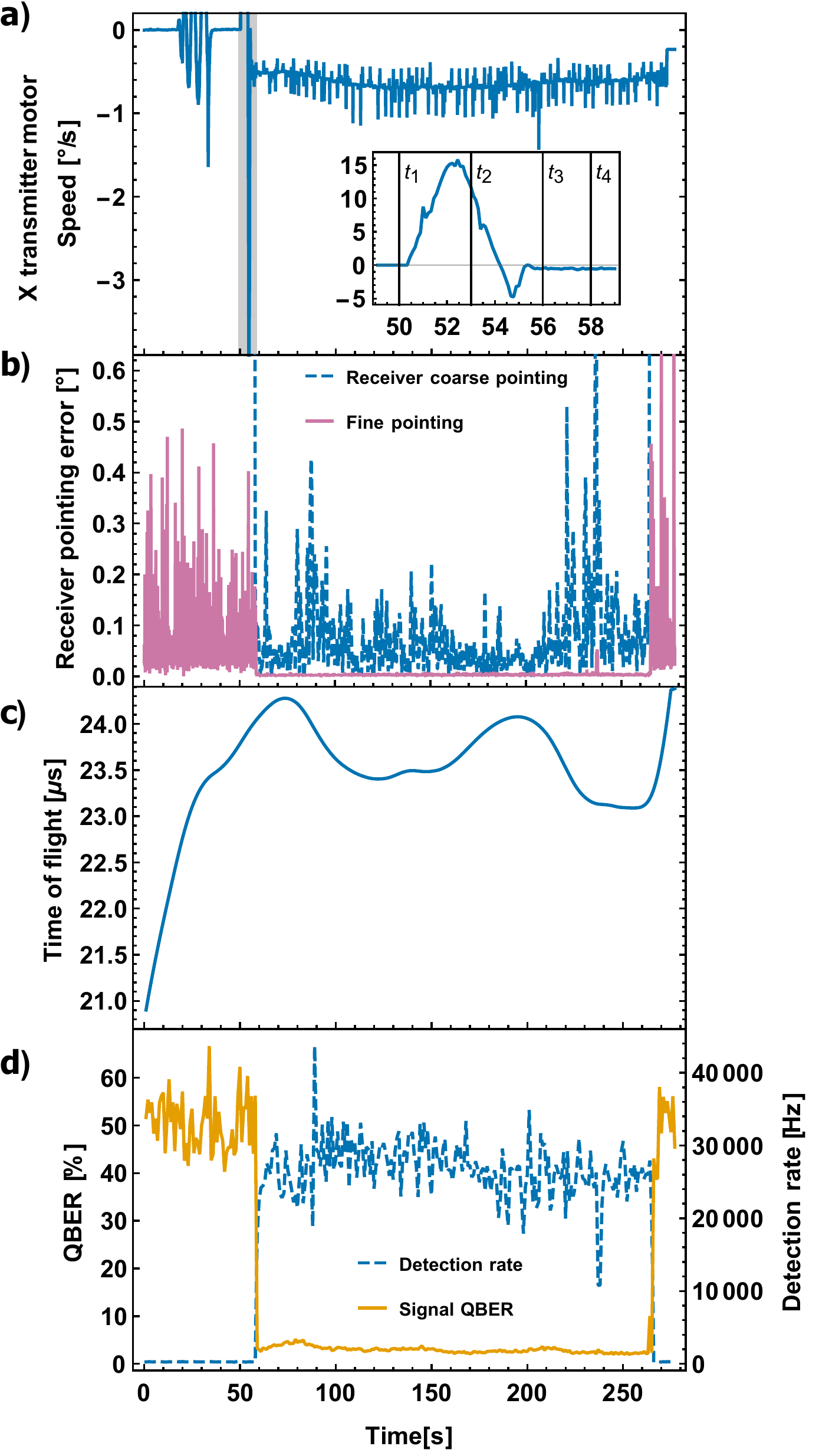}
		\end{subfigure}
		\hspace{0.02\linewidth}
		\begin{subfigure}[c]{0.475\linewidth}
			\centering\includegraphics[width=\linewidth]{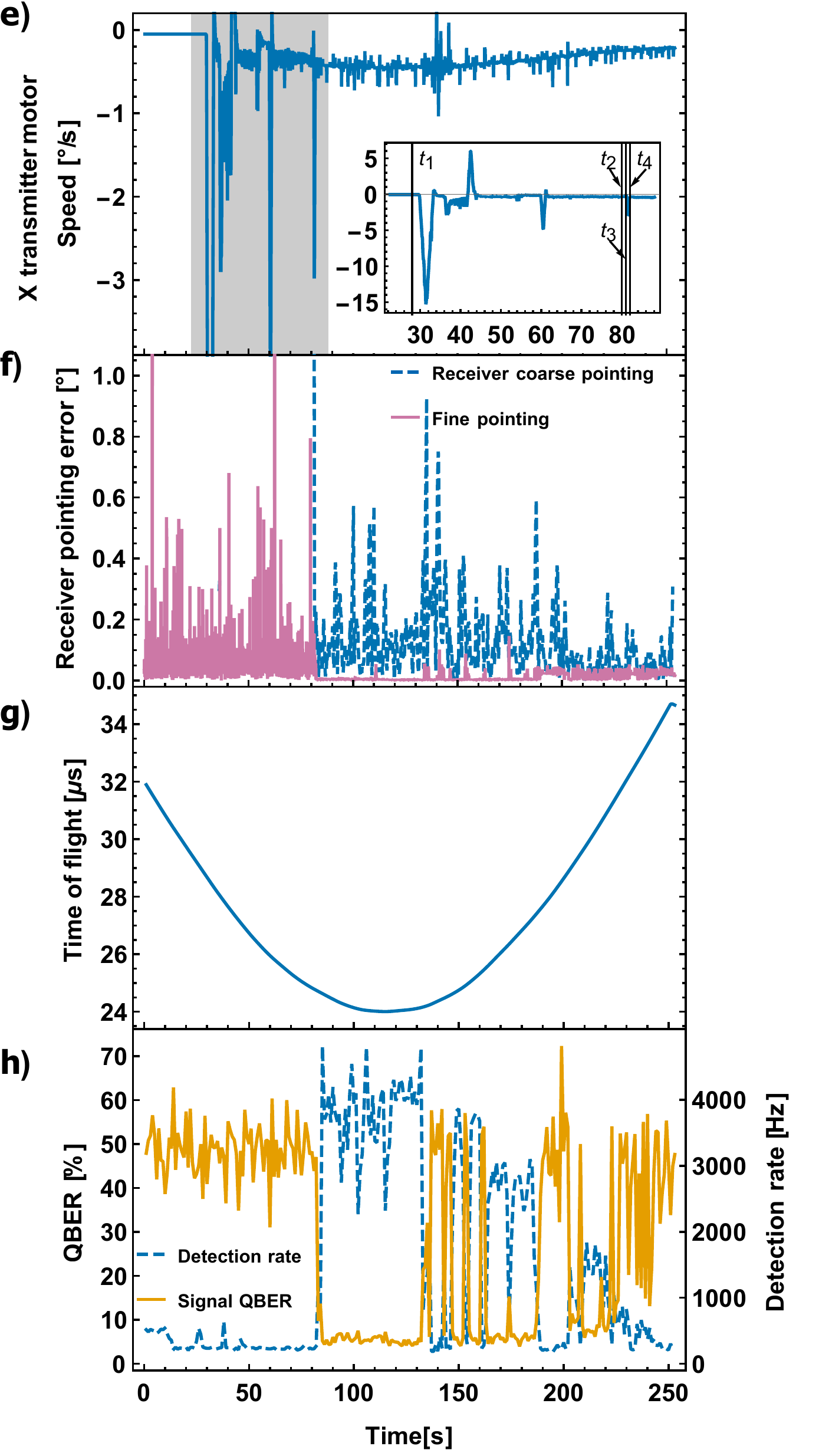}
		\end{subfigure}
		\caption{Results for the \SI{7}{\km} arc pass (left) and the \SI{7}{\km} straight-line pass (right).
		a) and e) show the speeds of the azimuthal (coarse) motor at the transmitter. The insets, corresponding to the shaded portions, show the motor speed during initial acquisition, with times $t_1$ through $t_4$ identifying establishment of the Wi-Fi link, identification of the beacon spot, lock to the beacon spot, and first counts received, respectively. The oscillation prior to this in a) is from a spiralling search state of the pointing software.
		b) and f) show coarse- and fine-pointing performance at the receiver. Where there are no coarse pointing data (e.g., at the beginning of a pass), no beacon spot was found in the camera image. This corresponds with large fluctuations in the fine-pointing deviation---in the absence of beacon light, the unit operates on electrical noise generated at the quad cell.
		c) and g) show the estimated time of flight of the photons from the transmitter to the receiver (used in event time-correlation), calculated from per-second GPS coordinates at each site. The smooth curve in g) is particularly characteristic of the straight-line pass, with a similar shape to that of a satellite pass.
		d) and e) show the total detection rates at the receiver and the QBER of the signal.}
		\label{fig:Pointing}
	\end{figure}

	Figure~\ref{fig:Pointing} shows observed results for two representative passes, including the motor speed of the transmitter in the horizontal axis and link acquisition stages, the coarse- and fine-pointing errors at the receiver, the calculated time of flight of the quantum signal from the transmitter to the receiver, the rate of detections of all four DM channels combined, and the quantum bit error rate (QBER) of the signal. There, the maximal angular speed was about \SIrange{0.5}{0.7}{\degree/\s} during beacon pointing lock, after initial acquisition.

	The mean measured loss of the quantum link during the flights varied from \SIrange{34.4}{51.1}{\dB}. Our theoretical loss model~\cite{BMH13} assumes a mid-latitude, rural atmospheric model in summer with the ground station located \SI{128}{\m} above sea level and \SI{5}{\km} visibility. Other model parameters include \SI{43}{\%} detector efficiency and receiver optical transmittance of \SI{59.7}{\%} (determined from the measured properties of the receiver prototype). We simulate the effect of atmospheric turbulence at our location using Hufnagel--Valley parameterization of atmospheric conditions~\cite{H74,H98}, with a sea-level turbulence strength of \SI{1.7e-14}{\m^{-2/3}} and high-altitude wind-speed of \SI{21}{\m/\s}. The measured pointing accuracy, aircraft altitude, and ground distance for each pass was also used. The divergence angle of the quantum beam could not be measured during the flight campaign. For the model we assume diffraction-limited divergence, resulting in lower bound theoretical loss estimates. Indeed, in the experiment, a number of passes were conducted with the transmitter intentionally slightly defocussed so as to avoid saturating the detectors. Consequently, the experimental losses we observed are generally higher than the theoretical losses. The difference between the theoretical loss of an arc pass and the minimum theoretical loss of a line pass at the same nominal distance is due to varying pointing accuracy experienced for each pass, as well as the actual ground distance and altitude deviating from nominal.

	For QKD analysis we utilize a signal-to-noise (SNR) filter~\cite{EHM12}, which assesses the total counts in each \SI{1}{\s} frame of data and discards any frame with counts less than a threshold, prior to distilling key bits. We choose thresholds between 1000 and 2500, depending on the pass. Background detection rates at the beginning and end of the pass are sufficiently low that those frames are discarded by the SNR filter. Some drop-outs can be seen in Fig.~\ref{fig:Pointing}h)---these frames are also discarded by the SNR filter.
	
	The source's intrinsic QBER, as predicted by the polarization correction system, varied between \SIrange[range-phrase={~and~}]{2.66}{5.08}{\percent} for each pass. At the receiver, the FPU, IOA, and fibers leading to the detectors were shielded with black cloth to minimize stray light entering the detectors, leading to typical total background detection rates of \SI{{\approx}285}{\Hz}.

	The QBER measured at the receiver drops to a few percent upon optical link lock, and rests at \SI{{\approx}50}{\percent} due to the random noise of background detections at all other times. For passes where secure key was generated, the QBER measured at the receiver, after the SNR filter, varied from \SIrange{2.96}{5.24}{\percent}. The received QBER during the first night flight was observed to be higher than for the second night, possibly due to an issue with the wave plate motorized stage controller.

	We generate secure key bits from the data collected during each pass using algorithms tailored for the asymmetric processing resources that would be available with a satellite platform \cite{BGH15}. These algorithms consist of source and receiver event time-correlation (performed at the ground station), error correction utilizing low-density parity check codes, and privacy amplification via reduced-Toeplitz-matrix two-universal hashes. To ensure security, the uncertainty due to the finite number of samples used to estimate link parameters must be taken into account. Of the six passes from which key could be extracted, five yielded secure key including these finite-size effects (where we use the common ten-standard-deviation heuristic to bound parameter estimates \cite{SLL09}). The remaining pass had too few counts and could only generate secure key assuming no finite-size effects.

	\section{Discussion and Conclusion}

	We have successfully demonstrated quantum key distribution to a satellite receiver payload prototype on an aircraft moving at up to \SI{259}{\km/\hour}. Our pointing and tracking system was able to establish and maintain an optical link with milli-degree precision over \SIrange{3}{10}{\km} distances while BB84 decoy-state signals were sent across the channel to the aircraft moving at the angular speeds of a LEO satellite. Our custom fine-pointing system, IOA, DM, and CDPU, along with the other commercial components, all performed in concert on the aircraft to generate secure keys, of tens to hundreds of kilobits in length, in various flight scenarios, including the straight-line paths approximating the apparent trajectory a LEO satellite. With source intrinsic QBER typically \SIrange{2}{4}{\percent} and post-processing algorithms representative of what would be achievable with a satellite platform, we extracted finite-size secure key for many of the tested passes.

	The details of path-to-flight modifications necessary to construct space-suitable versions of our receiver components varies. Some elements present on the CDPU daughterboard, for example, will need to be replaced with radiation-hard equivalent versions. Or, for the IOA, glues designed for low out-gassing must be used. Sensitivity of the Si-APDs in the DM to proton radiation in orbit is of particular note, as such radiation can significantly increase dark counts. However, strategies including cooling and thermal annealing~\cite{AHB17}, as well as laser annealing~\cite{LAH17}, are capable of mitigating these effects, and a space suitable prototype DM implementing these strategies is being developed.
	
	\begin{figure}[th]
		\centering
		\begin{subfigure}[c]{0.4\linewidth}
			\centering
			\includegraphics[width=\linewidth]{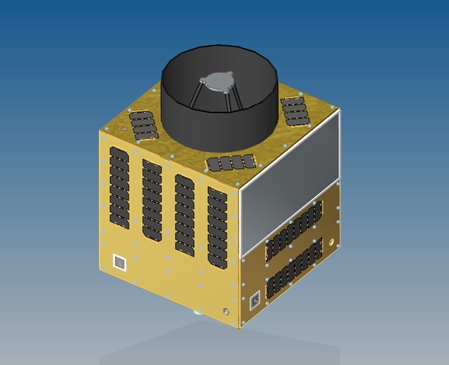}
		\end{subfigure}
		\hspace{0.025\linewidth}
		\begin{subfigure}[c]{0.25\linewidth}
			\centering
			\includegraphics[width=\linewidth]{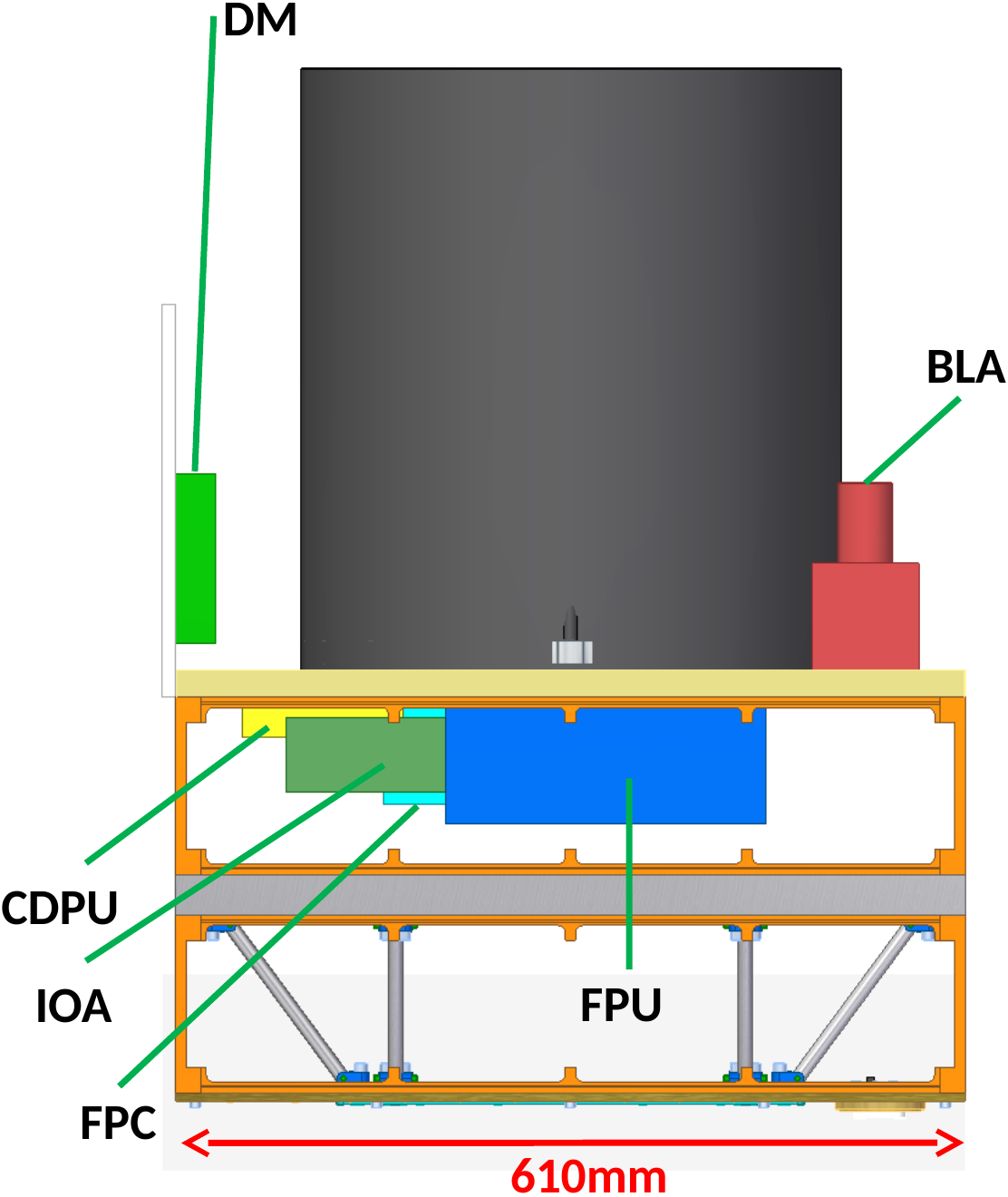}
		\end{subfigure}
		\caption{The UTIAS SFL NEMO-150 micro-satellite bus housing the quantum receiver payload. Left, external view showing extended telescope and baffle. Right, cross-section showing the possible placement of the main receiver payload components demonstrated during this airborne QKD campaign. Images provided by UTIAS SFL.}
		\label{fig:Sat}
	\end{figure}

	For pointing to a satellite from the ground, initial acquisition will likely not have a real-time classical communication link to exchange position data. In this case, however, predictions of the satellite position at the time when a link is to be established can be used, as the orbital trajectory of a satellite is predictable with far greater accuracy than the flight path of an airplane. In this context, point-ahead may be necessary (depending on the transmitter's divergence) to ensure that the quantum beam is coincident with the satellite when it arrives, owing to the satellite's motion during the time of flight of the optical signals. A fine-pointing system would likely also be required to achieve sufficient accuracy over the significantly larger transmission distance. For the aircraft, this was not necessary.

	One advantage of the uplink approach is source flexibility. While we have demonstrated only operation with a weak coherent pulse source here, we fully expect that QKD using entangled photon pairs generated at the appropriate wavelength by, for example, spontaneous parametric down-conversion will produce equivalent results under a BBM92-style protocol~\cite{BBM92}, with one photon of each pair measured on the ground. To support this, no aspect of the receiver prototype need be modified.

	Our system demonstrates the viability of an uplink QKD satellite mission. The core quantum components of a QKD satellite receiver have been demonstrated and have clear path to inclusion in space-faring system. In particular, see Fig.~\ref{fig:Sat} from a recent study conducted with UTIAS SFL, which shows our receiver hardware---FPU, FPC, IOA, DM, and CDPU---with minor modifications, cohesively integrated onto the flight-proven NEMO-150 micro-satellite bus. With the feasibility of performing uplink QKD with moving platforms well supported with satellite-ready hardware, QKD at the global scale utilizing satellite uplinks is within reach.

	\section{Acknowledgements}

	The authors acknowledge funding from the Canadian Space Agency Flights and Fieldwork for the Advancement of Science and Technology (FAST) program as well as the Space Technology Development Program (STDP), Ontario Research Fund, the National Sciences and Engineering Research Council, the Canadian Institute for Advanced Research, and the Canada Foundation for Innovation. CJP acknowledges support from the Natural Sciences and Engineering Research Council Canadian Graduate Scholarship--Doctoral Program and the Ontario Government Ontario Graduate Scholarship Program. JJ acknowledges support from the Korean Institute for Science and Technology. SK acknowledges support from the Mike and Ophelia Lazaridis Fellowship Program.
	 
	The authors thank Jeremy Dillon and the Flight Research Laboratory team at the National Research Council of Canada for their expertise in integrating and flying scientific aircraft payloads, training, and assistance during the flights. We thank the members of the Smiths Falls Flying Club, especially Peter Campbell, for access to their airfield and facilities, Phil Kaye for providing hangar space and assistance, and Ramy Tannous for assistance at the ground station. We thank Ian D'Souza and Jeff Kehoe for assistance with preliminary equipment test flights, and Rolf Horn for allowing us to set up a temporary ground station on his property. We also thank Dotfast-Consulting, Excelitas Technologies, Institut National d'Optique, Neptec Design Group, and Xiphos Systems Corporation for the development and support of the custom components.

CJP, EC, and TJ managed the project, and planned and executed logistics.
SK, CJP, and TJ conducted feasibility and aircraft flight-path studies, with link analysis conducted by CJP and JPB.
CJP, SK, JPB, BLH, and TJ designed and tested system components, with industry partners.
SK, EA, VM, and TJ designed and built the receiver detector module.
CJP and SK designed and assembled the receiver payload.
BLH developed the coarse pointing system, data acquisition, data processing, and polarization compensation system software, supervised by TJ.  
JPB, NS, and TJ designed and built the quantum source.
CJP, SK, JPB, JJ, SA, BLH, and TJ conducted outdoor full-system calibration and tests.
CJP and JJ integrated the receiver payload into the aircraft, assisted by BLH.
CJP, JPB, BLH and TJ developed and managed flight operations and mission tasking.
CJP operated the receiver in flight, assisted by JJ.
BLH conducted data acquisition and pointing at the ground station.
JPB operated the quantum source, assisted by BLH.
NS, SA, and TJ supported ground station operations.
CJP analyzed the data, supervised by BLH and TJ.
TJ conceived and supervised the project.
CJP and BLH wrote the manuscript, with contributions from all authors.

\end{document}